\documentclass[aps,prc,showpacs,superscriptaddress,onecolumn,12pt,floatfix,preprintnumbers,nofootinbib]{revtex4-2}

\usepackage{graphicx}

\makeatletter
\@ifclasswith{revtex4-2}{onecolumn}{\newcommand{\ifproofpre}[2]{#2}}{\newcommand{\ifproofpre}[2]{#1}}
\makeatother

\usepackage{amsmath}
\usepackage{dcolumn}
\usepackage{graphicx}
\usepackage{isotope}
\usepackage{hyperref}

\usepackage{xperiodspace}
\usepackage{liealg}
\usepackage{wignert}

\newcommand{\SI}[2]{\ensuremath{{#1}\,{#2}}}

\newcommand{\ang}[1]{\ensuremath{{#1}^{\circ}}}

\newcommand{\keV}{{\mathrm{keV}}}
\newcommand{\MeV}{{\mathrm{MeV}}}
\newcommand{\fm}{{\mathrm{fm}}}
\newcommand{\um}{{\mathrm{\mu m}}}
\newcommand{\mm}{{\mathrm{mm}}}
\newcommand{\cm}{{\mathrm{cm}}}

\newcommand{\uCi}{{\mathrm{\mu Ci}}}
\newcommand{\ns}{{\mathrm{ns}}}
\newcommand{\uA}{{\mathrm{\mu A}}}
\newcommand{\torr}{\mathrm{Torr}}
\newcommand{\tesla}{\mathrm{T}}

\newcommand{\percent}{\%}

\usepackage{xcolor}

\begin{document}
\ifproofpre{}{\count\footins = 1000}  

\title{Remeasuring the anomalously enhanced $B(E2; 2^+ \rightarrow 1^+)$ in $^8\mathrm{Li}$}

\author{S.~L.~Henderson}
\author{T.~Ahn}
\email{Corresponding author: tan.ahn@nd.edu}
\affiliation{Department of Physics and Astronomy, University of Notre Dame, 225 Nieuwland Science Hall, Notre Dame, Indiana 46556, USA}
\author{P.~J.~Fasano}
\affiliation{Department of Physics and Astronomy, University of Notre Dame, 225 Nieuwland Science Hall, Notre Dame, Indiana 46556, USA}
\author{A.~E.~McCoy}
\affiliation{Facility for Rare Isotope Beams, Michigan State University, MI 48824, USA}
\affiliation{Department of Physics, Washington University in Saint Louis, Saint Louis, MO 63130, USA }
\author{S.~Aguilar}
\author{D.~T.~Blankstein}
\author{L.~Caves}
\author{A.~C.~Dombos}
\affiliation{Department of Physics and Astronomy, University of Notre Dame, 225 Nieuwland Science Hall, Notre Dame, Indiana 46556, USA}
\author{R.~K.~Grzywacz}
\author{K.~L.~Jones}
\affiliation{Department of Physics and Astronomy, University of Tennessee, Knoxville, TN 37996, USA}

\author{S.~Jin}
\author{R.~Kelmar}
\author{J.~J.~Kolata}
\author{P.~D.~O'Malley}
\author{C.~S.~Reingold}
\author{A.~Simon}
\affiliation{Department of Physics and Astronomy, University of Notre Dame, 225 Nieuwland Science Hall, Notre Dame, Indiana 46556, USA}
\author{K.~Smith}
\affiliation{Department of Physics and Astronomy, University of Tennessee, Knoxville, TN 37996, USA}
\altaffiliation{Present address: Los Alamos National Laboratory, Los Alamos, NM 87545, USA}


\date{\today}


\begin{abstract}
The large reported $E2$ strength between the $2^+$ ground state and $1^+$ first
excited state of $\isotope[8]{Li}$, $B(E2; 2^+ \rightarrow 1^+)=
55(15)\,e^2\fm^4$, presents a puzzle.  Unlike in neighboring $A=7\text{--}9$
isotopes, where enhanced $E2$ strengths may be understood to arise from
deformation as rotational in-band transitions, the $2^+\rightarrow1^+$
transition in $^8$Li cannot be understood in any simple way as a rotational
in-band transition.  Moreover, the reported strength exceeds \textit{ab initio}
predictions by an order of magnitude.  In light of this discrepancy, we
revisited the Coulomb excitation measurement of this strength, now using
particle-$\gamma$ coincidences, yielding a revised $B(E2; 2^+ \rightarrow 1^+)$
of $19(^{+7}_{-6})(2)$~e$^2$fm$^4$.  We explore how this value compares to what might be
expected in the limits of rotational modesl.  While the present value is about a factor of three smaller than
previously reported, it remains anomalously enhanced.
\end{abstract}

\maketitle

\section{Introduction}
\label{sec:intro}

Large $E2$ transition strengths found in the $A=7-9$ mass
region~\cite{npa2002:005-007,npa2004:008-010} suggest that these nuclei are
significantly deformed, which gives rise to rotational structure as a dominant
feature in the low-lying
spectrum~\cite{inglis1953:p-shell,millener2001:light-nuclei}.  The strong
transitions between the ground state and first excited state in
$\isotope[7]{Li}$, $\isotope[7]{Be}$, and $\isotope[9]{Be}$, or between excited
states in $\isotope[8]{Be}$~\cite{datar2013:8be-radiative}, are interpreted as in-band
rotational transitions.  This deformation, in turn, is understood to arise from
cluster molecular structure~\cite{vonoertzen1996:be-molecular,*vonoertzen1997:be-alpha-rotational,freer2007:cluster-structures,kanadaenyo2012:amd-cluster,maris2012:mfdn-ccp11,kravvaris2017:ab-initio-cluster-8be-10be-12c}, \textit{e.g.}, with $\isotope[7]{Be}$ as a
$\isotope[3]{He}+\isotope[4]{He}$ dimer, and the heavier $\isotope{Be}$ isotopes
as $\isotope[4]{He}+\isotope[4]{He}$ plus neutrons.

In general, $E2$ strengths provide a probe of nuclear structure and its
evolution~\cite{casten2000:ns}.  For light $p$-shell nuclei, which are
accessible to \textit{ab initio} nuclear theory by a variety of approaches,
 \textit{ab initio} calculations can provide qualitative insight into the
structural origin of the $E2$ strengths \cite{wiringa2000:gfmc-a8,pervin2007:qmc-matrix-elements-a6-7,pastore2013:qmc-em-alt9,pastore2014:qmc-em-8be}. Meanwhile,
experimental measurements can provide quantitative validation of the ability of
calculations to faithfully describe the nuclear
system~\cite{henderson2019:7be-coulex}.

Taken in this light, the large reported $E2$ strength between the $2^+$ ground
state and $1^+$ first excited state of $\isotope[8]{Li}$, $B(E2; 2^+ \rightarrow
1^+)= 55(15)\,e^2\fm^4$~\cite{brown1991:8li-coulex}, presents a puzzle.  This strength
corresponds to $\sim58$ Weisskopf units, which would be considered collective
even in much heavier mass regions.
The $M1$ strength for the $2^+_1 \rightarrow 1^+_1$ transition can be deduced from the measured lifetime of this transition due to the dominance of the $M1$ component. The measured lifetime of the $1^+_1$ level was measured previously by the Doppler-shift attenuation method \cite{Throop1971,COSTA1972}. The large value of the corresponding $M1$ transition strength is interesting in its own right, but we focus on the $E2$ component, which give complementary information to the $M1$ transition strength. As the $2^+_1 \rightarrow 1^+_1$ transition in $^8$Li is dominated by the $M1$ component, we have chosen to use the technique of Coulomb excitation to selectively measure the $E2$ transition strength.

It is not unexpected that $\isotope[8]{Li}$ would be deformed.  For
instance, the ground state of the mirror nuclide $\isotope[8]{B}$ is suggested to
have proton halo structure~\cite{minamisono1992:8b-quadrupole-beta-nmr,smedberg1999:8b-fragmentation-halo,jonson2004:light-dripline,korolev2018:8b-pscatt-halo} as a deformed $\isotope[7]{Be}$
core, with a loosely-bound proton in a spatially-extended molecular
orbital~\cite{henninger2015:8b-fmd}.  The ground-state spectroscopic quadrupole
moment of $\isotope[8]{Li}$ is similar in magnitude to those of its deformed
neighbors~\cite{stone2016:e2-moments}.

Nonetheless, an enhanced $2^+\rightarrow 1^+$ transition cannot be easily
understood as an in-band rotational transition, like the transitions in
neighboring nuclei.  As least in a conventional axially-symmetric rotational
picture~\cite{bohr1998:v1,rowe2010:collective-motion}, if the $2^+$ ground state
is the band head of a $K=2$ band, there is no $1^+$ band member, and the
transition to the $1^+$ excited state is at most an interband transition
(between $K=2$ and $K=1$ band heads).

Moreover the reported strength exceeds \textit{ab initio} Green's function Monte
Carlo (GFMC) predictions~\cite{pastore2013:qmc-em-alt9} by nearly two orders of
magnitude, despite the same calculations predicting a result for the quadrupole moment of the $2^+$ ground state, which is in line with
experiment~\cite{stone2016:e2-moments}.  In this paper, we present \textit{ab initio} no-core
shell model (NCSM) calculations of the type presented in Ref.~\cite{barrett2013:ncsm}, with various
interactions, which are likewise inconsistent with the reported $E2$
enhancement.  Although the absolute scale of $E2$ strengths is poorly convergent
in NCSM calculations, robust and meaningful predictions may be obtained by
calibration~\cite{calci2016:observable-correlations-chiral} to the
experimentally-known electric quadrupole moment.

The large reported $E2$ strength in $\isotope[8]{Li}$~\cite{brown1991:8li-coulex} was
measured by Coulomb excitation with a radioactive beam of $\isotope[8]{Li}$,
where the inelastically-scattered $\isotope[8]{Li}$ nuclei were detected by measuring their energy using a
$\isotope{Si}$ detector.  Such an experiment is ostensibly susceptible to
events coming from $^8$Li that is produced in its excited state in the primary reaction rather than those coming from the Coulomb excitation of $^8$Li in the secondary target, which would result in an inflated measured
Coulomb-excitation cross section and thus the extracted $E2$ strength.  To eliminate such a
possible source of error, we revisit this radioactive-beam Coulomb-excitation
measurement, but now with gamma-ray detection capability, to impose a
coincidence requirement between the detection of the inelastically-scattered
$\isotope[8]{Li}$ nucleus and the $1^+\rightarrow 2^+$ deexcitation gamma ray.
We use an array of high-efficiency $\isotope{La}\isotope{Br}_3$ gamma-ray
detectors, in coincidence with a $\isotope{Si}$ particle detector centered on the secondary target.  While we measure a smaller value than
previously reported, the change is not sufficient to bring experiment in line
with current theoretical understanding of the structure.

We first outline the present radioactive beam experiment with the \textit{TwinSol}
low-energy radioactive nuclear beam apparatus at the University of Notre Dame Nuclear
Science Laboratory~\cite{Becchetti2003} (Sec.~\ref{sec:expt}) and detail the
subsequent analysis used to extract the $2^+\rightarrow1^+$ strength from
Coulomb excitation (Sec.~\ref{sec:analysis}), including an assessment of
two-step and other possible contributions.  We then discuss this strength  in the
context of rotational and \textit{ab initio}
 descriptions (Sec.~\ref{sec:discussion}). These results were reported
in part in Ref.~\cite{henderson2021:thesis}.

\section{Experiment}
\label{sec:expt}

In order to produce a beam of \isotope[8]{Li}, the 10 MV Tandem Van De Graaff at
the University of Notre Dame Nuclear Science Laboratory (NSL) was used to accelerate
a \SI{4.5}{e\uA} beam of \isotope[7]{Li} $3^+$ ions to \SI{26}{\MeV}. This beam was steered
to a production gas cell, which had a \SI{12}{\um} thick \isotope[9]{Be} foil
placed on the downstream side of the gas cell to serve as the production
target. The gas cell was filled with \SI{300}{\torr} of He gas to help cool the
\isotope[9]{Be} target and a \SI{4}{\um} Ti foil on the upstream side of the gas
cell was used to contain the gas. The \isotope[8]{Li} beam was produced by the
\isotope[9]{Be}(\isotope[7]{Li},\isotope[8]{Be})\isotope[8]{Li} reaction at an
energy of \SI{22.7(5)}{\MeV} (80\% of the Coulomb barrier), along with other isotopes produced by competing
reactions. This cocktail beam was sent into the \textit{TwinSol} apparatus
\cite{Becchetti2003}. \textit{TwinSol} consists of two superconducting solenoid
magnets that are used as magnetic lenses to focus the radioactive beam of interest and
eliminate contaminants. More details on how the \textit{TwinSol} apparatus was
used in this experiment are given in Ref.~\cite{henderson2019:7be-coulex}. The first of these two solenoids was set to
\SI{2.43}{\tesla} to best focus the \isotope[8]{Li} beam through a \SI{9}{\mm}
diameter collimator placed between the solenoids, which eliminated the majority
of the contaminants. The second solenoid was set to \SI{1.40}{\tesla}, to refocus
the \isotope[8]{Li} beam after it had passed through the collimator.

After exiting \textit{TwinSol}, the \isotope[8]{Li} beam was sent into our
scattering chamber. The first element the beam encountered was a \SI{9}{\mm}
radius collimator, to eliminate divergent aspects of the beam and define the
beam spot. The collimator and the experimental setup is shown in Fig.~\ref{fig:detector_setup}.
Directly downstream of the collimator was a target ladder with a
\SI{1}{\um} thick \isotope[197]{Au} target, an empty frame, and a Si surface
barrier detector for determining beam purity. During beam development, the beam was sent onto the surface barrier detector for
particle identification and the final beam components were identified as being \SI{98}{\percent}
\isotope[8]{Li} with \isotope[7]{Be} and scattered
\isotope[7]{Li} making up the majority of the contaminants. The optimized radioactive beam was sent onto the the gold foil for the
duration of the experiment. Placed \SI{34.6}{\mm} downstream from the gold foil,
a \SI{1000}{\um}-thick S7 annular Si detector, made by Micron Semiconductor
Limited \cite{micron}, was used to measured the beam that scattered from the \isotope[197]{Au}
target in the angular range of \ang{20.6} - \ang{45.3} degrees. The upstream
side of the Si detector is segmented into 45 \SI{0.5}{\mm} concentric
rings from its \SI{13}{\mm} inner radius to its \SI{35}{\mm} outer radius while
the downstream side of the detector is segmented into 16 radial sectors of
\ang{22.5} each, giving precise radial positions and the ability to measure beam offsets. Due to the limited number of electronic channels available, the ring
electronic channels were combined in pairs to make 22 rings, each effectively
\SI{1}{\mm} wide.  Additional beam parameters that were deduced from the measurement of the beam with the Si detector is given in Sec.~\ref{sec:analysis}.

Outside of the scattering chamber, 10 LaBr$_3$ detectors with
\SI{2}{\mathrm{in.}} $\times$ \SI{2}{\mathrm{in.}} cylindrical crystals from the
HAGRiD array \cite{Smith2018} were placed \SI{17.2}{\cm} away from the target,
in order to measure $\gamma$ rays emitted by Coulomb-excited \isotope[8]{Li}
nuclei.  The arrangement of the LaBr$_3$ detectors can be seen in
Fig. \ref{fig:detector_setup}. The detectors were placed symmetrically around
the chamber at \ang{30}, \ang{60}, \ang{90}, \ang{120}, and \ang{150} on both
sides of the beam axis. The $\gamma$-ray peaks from the intrinsic radiation of the LaBr$_3$ detectors, $\beta$ decay of
\isotope[138]{La}, were used to gain match the different detectors
before the experiment and were monitored throughout the experiment.

 \begin{figure}
 \centering
 \includegraphics[width=\ifproofpre{1.0}{0.5}\columnwidth]{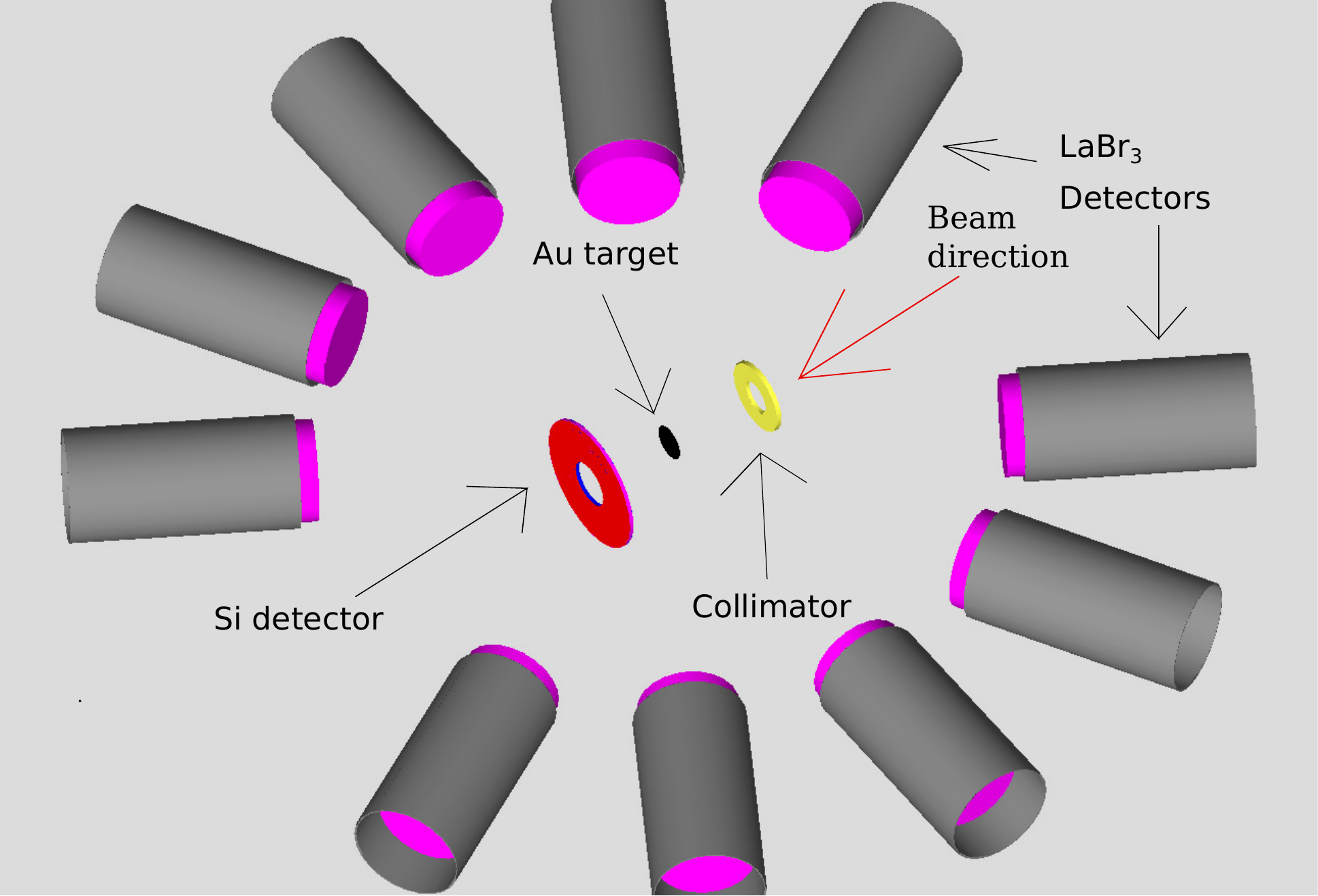}
 \caption{(Color online) The experimental setup is shown. Ten LaBr$_3$ detectors
   from the HAGRiD array are placed at \ang{30}, \ang{60}, \ang{90}, \ang{120},
   and \ang{150} with respect to the beam axis surrounding the Au
   target, a brass collimator, and a S7 Si detector inside the scattering chamber
   (not shown).}
 \label{fig:detector_setup}
 \end{figure}

The signals from the Si and LaBr$_3$ detectors were run through preamplifiers
and into a digital data acquisition (DAQ) system using Pixie-16 modules from
XIA, LLC \cite{xia}. In this experiment, a hit in any detector channel defined an event in
the DAQ, with a hit in any another detector channel being considered coincident and packaged
together into the same event if it occured within a \SI{500}{\ns} time window of
the original event. In the second half of the experiment, this timing window was
reduced to \SI{100}{\ns}, to reduce the number of random coincidences in
the $\gamma$-ray spectrum. The experiment was run for a total of 5 days with a beam rate of $4.1(3) \times 10^5$ pps. The precise determination of the beam rate is discussed in Section \ref{sec:analysis}.

At the end of the experiment, a \SI{1.44}{\uCi} \isotope[152]{Eu} source was
placed at the target location and the multiple $\gamma$ rays from its decay were
used to calibrate each LaBr$_3$ detector in energy and determine its
$\gamma$-ray efficiency. The entire array was found to have a total $\gamma$-ray
efficiency of \SI{0.82}{\percent} at \SI{1}{\MeV}. The energy resolutions of
the LaBr$_3$ detectors in the array were \SI{1.6}-\SI{2}{\percent} at
\SI{1408}{\keV}. These resolutions were sufficient to cleanly resolve
our $\gamma$-ray of interest (981 keV) since this energy is far enough away from the background $\gamma$-rays
seen in the Doppler uncorrected spectrum.

\section{Analysis}
\label{sec:analysis}

To determine the $E2$ transition strength from our experimental observables, we needed to precisely determine our integrated beam rate over the course of the experiment and determine the $\gamma$-ray yield from the Coulomb-excited $^8$Li nuclei. The details of each step is outlined below.

In order to determine the total,
integrated \isotope[8]{Li} beam current,
we simulated the process of producing an in-flight beam
with \textit{TwinSol}, which leads to an extended spot size on the target.
We modelled the beam as having a finite radius and offset from the center of the target. A
fixed collimator upstream of the gold foils restricted the beam spot to a 9 mm
radius, which was chosen to match the size of our gold target. Due to the
proximity of the target to the Si detector, a diffuse beam will cause the
\isotope[8]{Li} ions seen in the rings of the Si detector to come from a range of
scattering angles. The Si detector also shows some up-down asymmetry in the
measured rates in the sectors, indicating that the beam was offset to some
degree. Because of the diffuse and asymmetric nature of the beam, comparing the
distribution of the particles in the rings of the detector to a Rutherford
distribution does not yield an accurate fit. We used a Geant4
\cite{Agostinelli2003,Allison2006,Allison2016} simulation to model the width and
offset of the beam on target, accounting for the various scattering angles that result in counts in a single Si detector ring. The
distribution of scattering angles seen in one of the inner rings of the Si detector is shown in Fig.~\ref{fig:AngleDistribution}.
\begin{figure}
\centering
\includegraphics[width=\ifproofpre{1.0}{0.5}\columnwidth]{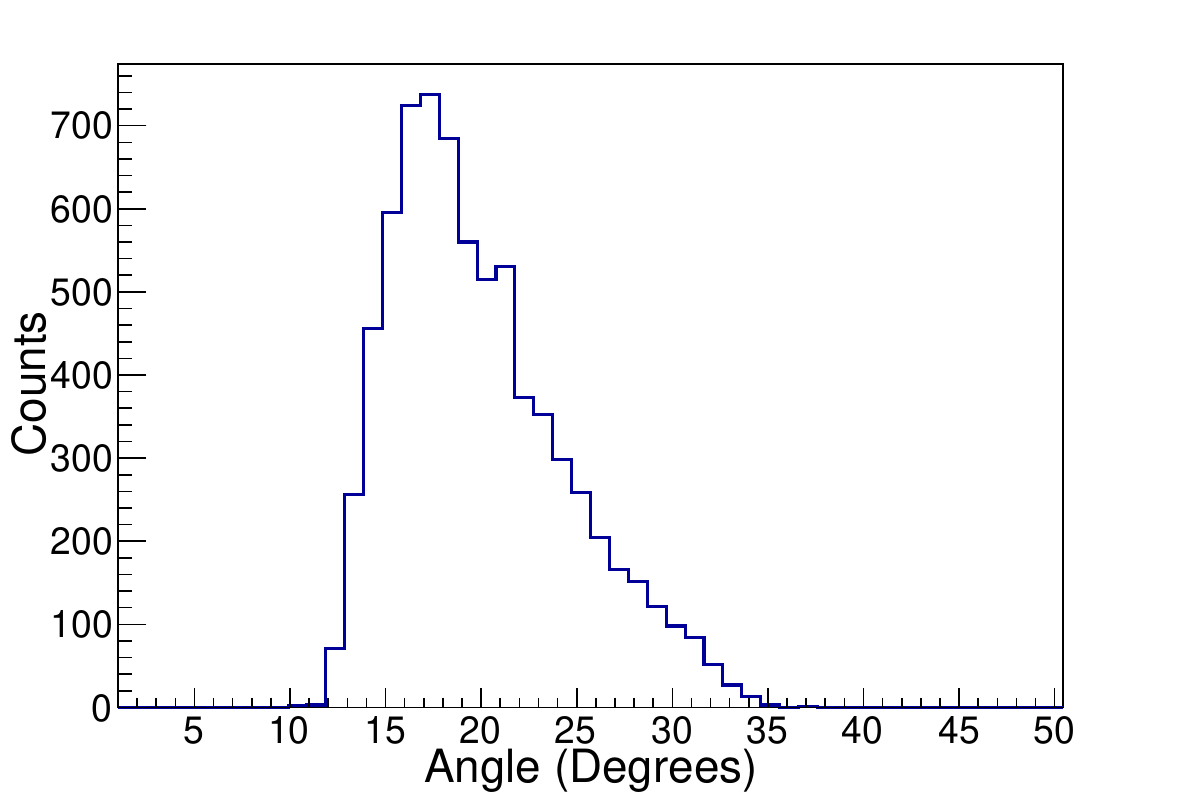}
\caption{(Color online) A simulation of the distribution of angles of
  \isotope[8]{Li} particles observed in the first ring of the Si detector used in our
  analysis when using the beam spot determined by our simulation. The number of
  particles seen here is not representative of the total particles seen during
  the experiment, though the shape of the distribution should be accurate.}
\label{fig:AngleDistribution}
\end{figure}
The radius of the beam and its offset
from the beam axis were varied in the simulation over a range of values
(1-7 mm for the radius and 1-4 mm for the
offset) to reproduce the distribution seen in both the rings and sectors.  We found a \SI{7}{\mm} beam radius
and a \SI{1}{\mm} offset best reproduced the shape of the Si detector ring and
sector data. The data seen in the Si rings and the simulation results are shown in Fig.~\ref{fig:particleSpectrum}. It can be seen that the simulation reproduces the shape of the measured counts in each Si detector ring very well. After the simulation reproduced the shape of the experimental data, it was scaled to have the
same magnitude as the total number of counts measured during the experiment and a beam rate of $4.1(3) \times 10^5$~pps was extracted from the
scaling. The beam rate uncertainty was estimated by changing the beam parameters
and the beam scaling until the simulation results exceeded the experimental
uncertainties.
\begin{figure}
\centering
\includegraphics[width=\ifproofpre{1.0}{0.5}\columnwidth]{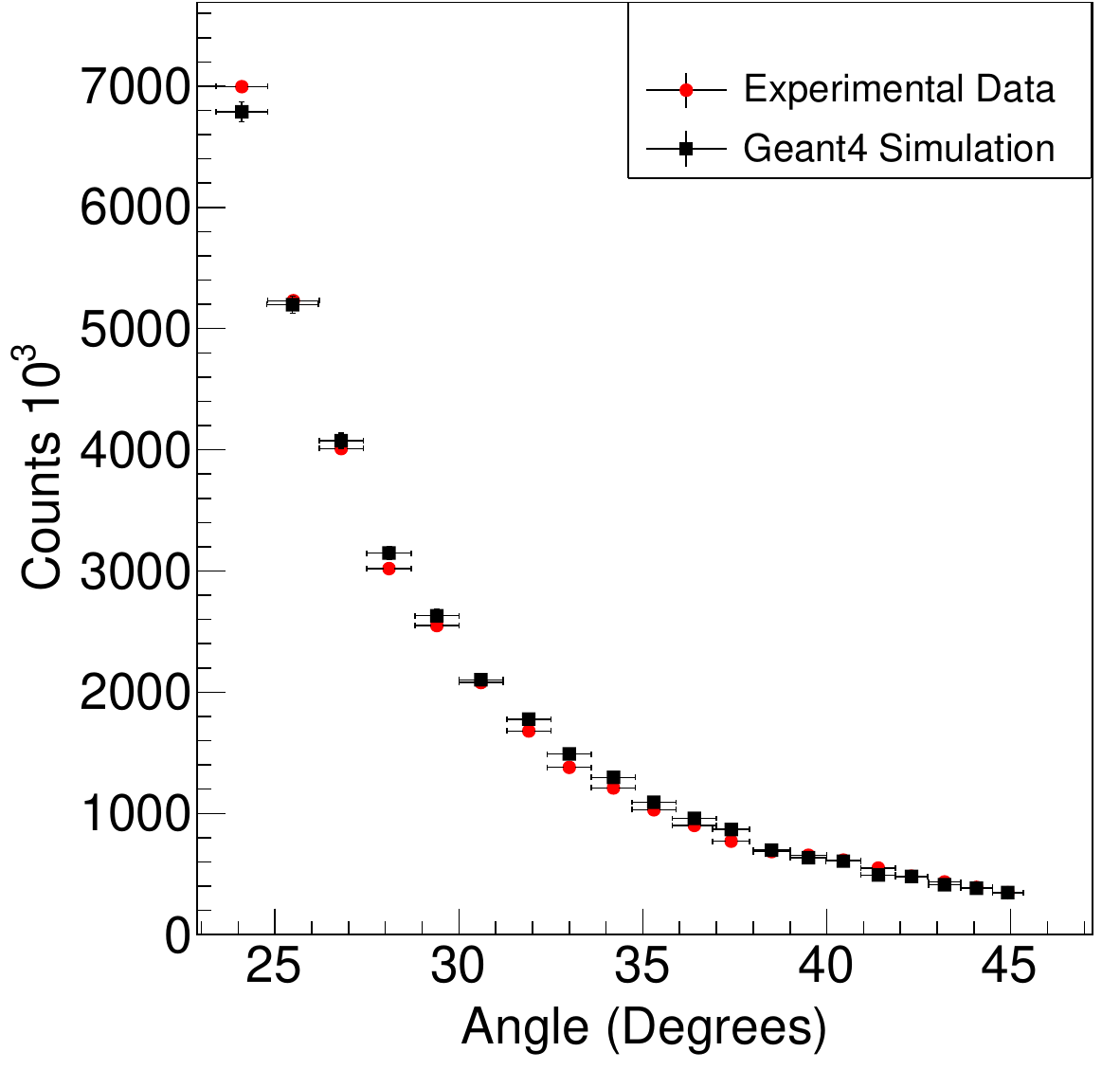}
\caption{(Color online) A plot of \isotope[8]{Li} ions measured in the rings of
  the Si detector (circles) and the Geant4 simulated data (squares). The angles
  represent the center of each ring of the Si detector with the horizontal
  uncertainties showing the full angular extent of each ring.}
\label{fig:particleSpectrum}
\end{figure}

A coincidence gate was placed on the scattered \isotope[8]{Li} peak seen
in our Si detector in
order to reduce the number of $\gamma$ rays from room background and
the intrinsic radioactivity of the LaBr$_3$ detectors. As the Si ring and sector and
LaBr$_3$ detector responsible for the coincidence event were known, the
geometric angle between the scattered \isotope[8]{Li} particle and the emitted
$\gamma$ ray was determined and used to correct for the Doppler shift of the
$\gamma$-ray energy. During the analysis, a large amount of beam background was observed in the two innermost rings of the Si detector, which
were therefore not included in this analysis.

\begin{figure}
\centering
\includegraphics[width=\ifproofpre{1.0}{0.6}\columnwidth]{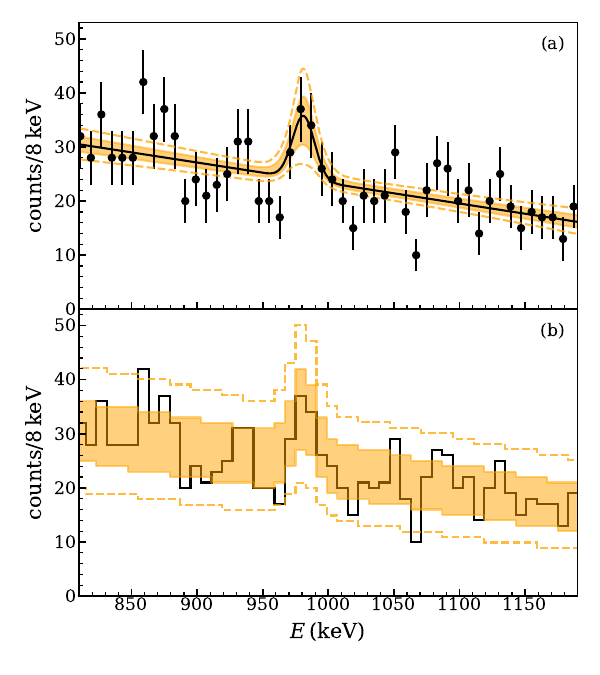}
\caption{(Color online) The Doppler-corrected $\gamma$-ray spectrum summed over
  the entire experiment and in coincidence with the scattered-particle
  peak in the silicon detector. The spectrum is binned to 8
  keV/bin. The $\gamma$-ray peak corresponding to the $2^+ \rightarrow 1^+$
  transition of \isotope[8]{Li} is seen at \SI{981}{\keV}. In panel (a) the solid curve of the Gaussian function on a linear background with the most likely parameters is overlaid on the data, with the error bands representing 68\% and 95\% confidence intervals on the line of best fit. In panel (b) the uncertainty in the fit is folded with the Poisson statistical error and overlaid on the data; the spectrum is expected to lie within the band with 68\% and 95\% confidence.}
\label{fig:gamma_spectrum_fit}
\end{figure}
Additionally, some of the random coincidences seen in the spectrum were
eliminated by requiring a tight time coincidence between the LaBr and Si
detector signals. The original coincidence window used in the experiment was 500 ns wide,
but an additional timing gate was added in the offline analysis. The timing offsets between different rings of
the Si detector and different LaBr$_3$ detectors were aligned and a 30 ns timing
gate was placed over the position of the particle-$\gamma$ coincidences in the time difference spectrum. As the number of particle-$\gamma$ coincidences were low, this position was determined by observing the Doppler-corrected $\gamma$-ray spectrum while using a moving gate in the time-difference spectrum. The position and width were chosen to provide a robust $\gamma$-ray peak of interest at the expected peak position of \SI{980.8(1)}{\keV}, the known literature value of the the $1^+ \rightarrow 2^+$ transition \cite{COSTA1972}.
The $\gamma$-ray spectrum in this energy range is shown in Fig.~\ref{fig:gamma_spectrum_fit} and a small peak is visible above the background at the expected energy.

In order to determine the number of counts in the peak corresponding to the 981-keV $1^+ \rightarrow 2^+$ transition in a robust way, we used Bayesian inference in modeling the spectrum with a Gaussian peak on a linear background:
\begin{equation}
n(E) = n_0 + m(E-E_0) + w\frac{A}{\sqrt{2\pi s^2}} e^{-(E-E_0)^2/2s^2},
\end{equation}
where $n_0$ and $m$ characterize the amplitude and energy slope of the background, $w$ is the width of the energy bins, and $A$, $s$, and $E_0$ characterize the area, width, and centroid energy of the peak, respectively. We calculated the posterior distribution of the fit parameters using a Markov-Chain Monte Carlo (MCMC) algorithm, implemented with the \texttt{emcee} package \cite{foremanmackey2013:emcee}. The likelihood function was obtained assuming that the probability for a datum in our spectrum is given by the Poisson distribution. Uniform priors were used for all parameters except the position and width of the peak. The uniform priors were properly normalized for a range of 0 to 75 for $n_0$, 0 to -1 for $m$, and 0 to 65 for the peak area $A$. For the peak position, a Gaussian prior with a $\sigma$ width corresponding to 1.5 keV was used. For the peak width~$s$, the prior probability was given as $P(s) = \frac{\alpha^{\alpha-1}}{s_0^2 \Gamma(\alpha-1)}e^{-\alpha \left(s_0/s\right)^2} \left(\frac{s_0}{s}\right)^{2\alpha}$, where $s_0=10\,\mathrm{keV}$ is the most likely or ``expected'' width of the peak, and the $\alpha=3$ specifies the shape of the distribution of possible widths. The value of $s_0=10\,\mathrm{keV}$ was chosen based on the measured widths of the LaBr$_3$ detectors' intrinsic resolution from calibration data. Two-dimensional marginalized distributions of the calculated posterior probability density for all pairs of parameters and the marginalized distributions of one parameter are shown in Fig.~\ref{fig:posterior_pdfs}.
\begin{figure*}[p]
\centering
\includegraphics[width=1.0\textwidth]{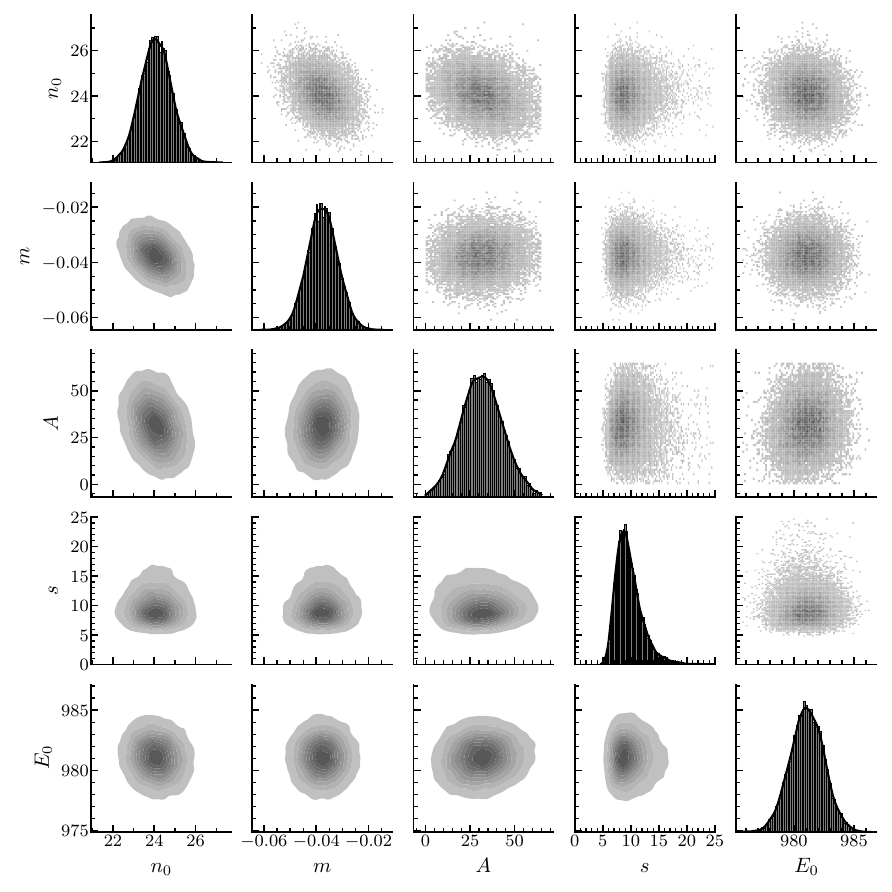}
\caption{The marginalized distributions for the calculated posterior probability density function of the Gaussian and linear background parameters. The two-dimensional contour plots show the marginalized distribution for each pair of parameters and the one-dimensional histograms show the marginalized distributions for each parameter along the horizontal axis. The plots in the lower triangle are smoothed using kernel density estimation from the package \texttt{seaborn} \cite{waskom2021:seaborn}, while the plots in the upper triangle are histograms of the Monte Carlo sample points. Similarly, the solid curves in the diagonal plots are smoothed using kernel density estimation, while the histograms represent the sample points.}
\label{fig:posterior_pdfs}
\end{figure*}

The curve corresponding to the most likely parameters is plotted over the $\gamma$-ray spectrum in Fig.~\ref{fig:gamma_spectrum_fit}(a), along with 68\% and 95\% confidence intervals on the fit curve. In Fig.~\ref{fig:gamma_spectrum_fit}(b), we show the effect of including the Poisson statistics into the model prediction. From the posterior probability distribution for the model parameters (generated by the MCMC algorithm), we generate 15,000 synthetic $\gamma$-ray spectra; 68\% of the generated spectra fall within the shaded band in Fig.~\ref{fig:gamma_spectrum_fit}(b), while 95\% of the generated spectra fall between the dotted lines. The fact that the band in Fig.~\ref{fig:gamma_spectrum_fit}(b) is substantially wider than the band in Fig.~\ref{fig:gamma_spectrum_fit}(a) reflects the fact that, even for reasonably well-constrained fit parameters, we expect significant scatter in the spectrum entirely because of low counting statistics.

The most likely value of the area of the Gaussian peak is
$31^{+13}_{-11}$ counts,
where the error bar comes from the limits that correspond to 1$\sigma$ normal probabilities.
For the other parameters, the maximum likelihood is given by $s = 8.4^{+2.5}_{-1.7}$, $E_0 = 981.1(14)$, $m = 0.038(6)$, and $n_0 = 24.1(7)$, again with 1$\sigma$ uncertainties.
It is worth noting that the marginal distribution for the peak area is not Gaussian, especially for values approaching zero. Based on the area parameter's marginal distribution,
the interval corresponding to $2\sigma$ (97.7\% probability) excludes a lower value below 9.0 counts.
The implication of these peak area values on the deduced $B(E2)$ values will be discussed later in this section.

In addition to looking at the posterior distributions, we have also analyzed the $\gamma$-ray spectrum using a Bayes factor analysis. The Bayes factor is a ratio of posterior probabilities for two hypotheses  and can be used in model selection \cite{Kass1995}. One of these hypotheses can be the null hypothesis and the one we have chosen is that there is no peak that exists at 981 keV. In this hypothesis, the spectrum is described only by a linear background. The Bayes factor disfavors the use of many adjustable parameters so a hypothesis with a smaller number of parameters that can describe the data equally well will be favored over one with a larger number of parameters. This means that model selection using the Bayes factor is not solely based on how well the data is fit by the model but will favor the simpler model in the case that both models fit the data equally well. One of the most important inputs into the Bayes factor is the prior probability for the peak area. As mentioned previously, we have chosen a uniform prior for the area with a range of 0 to 65 counts, a reasonable and generous assumption. In order to calculate the Bayes factor, we use a sequential Monte Carlo (SMC) algorithm, as implemented by the \texttt{PyMC3} package~\cite{Salvatier2016}. We obtain a Bayes factor of 9.0 when comparing the ``peak'' hypothesis to the null hypothesis. This value falls squarely in the range given by Ref.~\cite{Kass1995} of 3 to 20, which is considered ``positive'' evidence against the null hypothesis.  Although this value of the Bayes factor does not completely exclude the null hypothesis,  there is substantial evidence for the ``peak'' hypothesis such that it should not be ruled out.

Using the measured efficiency of the LaBr$_3$ array and the most likely value for the number of counts observed in the 981 keV $\gamma$-ray peak, we were able to determine the absolute $\gamma$-ray yield for the experiment.

The final step was to determine the \isotope[8]{Li} $B(E2; 2^+ \rightarrow 1^+)$ value from the
total $\gamma$-ray yield in the experiment. We used a version of the Winther-De
Boer Coulomb excitation code, which is based on the semi-classical theory of
Coulomb excitation \cite{Alder1956}. This version of the code was capable of
calculating electric dipole to hexadecapole transitions and also multiple excitations. After inputting an ${E}2$ matrix element, the code
outputs a differential cross section which can be used to match the output of
the $\gamma$-ray yield measured in the experiment. Due to the shape of the
\isotope[8]{Li} beam, varying scattering angles in each Si detector ring, and Si detector geometric efficiencies, it would be difficult to account for in a simple
calculation. Therefore, we used the Geant4 simulation to use the event-by-event simulated angle to
determine the proper Coulomb-excitation probability given by the Winther-De Boer calculations. We determine the $B(E2)$ by finding the value that matches our experimental $\gamma$-ray yield through the Geant4 simulation.
The \isotope[8]{Li}
$B(E2; 2^+ \rightarrow 1^+)$ value we obtained is $\SI{19^{+7}_{-6}}{e^2\fm^4}$, which includes the
$\gamma$-ray yield statistical and beam rate uncertainties.

Coming back to the values of the peak area that are excluded by
a $2\sigma$ 
interval, the $B(E2; 2^+ \rightarrow 1^+)$ value less than
\SI{5.2}{e^2\fm^4} 
 are excluded with a
97.7\%
probability based on our data. This probability does not favor theoretical predictions, which will be discussed in detail in Sec.~\ref{sec:discussion}.

The two sources of systematic uncertainties are the $M1$ component of the
excitation probability and contributions from electric dipole polarizability, i.e., virtual $E1$ excitations to  collective structures at high energy.
Based on a study of \isotope[7]{Li} \cite{Hausser1973},
which should be a fairly close analogue to \isotope[8]{Li},  we estimate the $M1$
contribution at the forward angles in our experiment to be between 2-3\percent. Much
less easily understood is the effect of the $E1$ dipole polarizability. Unlike \isotope[7]{Li}, which has a strong virtual excitation to the
break up of \isotope[3]{H}+$\alpha$, \isotope[8]{Li} mainly virtually excites to
levels that decay by neutron emission. Studies on the $E1$ dipole polarization effect in $^6$Li and $^7$Li suggest that this effect is small, on the order of less
than 10\% at forward angles \cite{Hausser1973,Disdier1971}. With a conservative estimate of 3$\percent$ uncertainty due to the M1 excitation and a 10$\percent$
uncertainty due to the effect of $E1$ dipole polarizability, we account for these by assigning a systematic uncertainty in the $\gamma$-ray peak area, which results in an uncertainty in the $B(E2)$ of 2 e$^2$fm$^4$. More precise
measurements in the future that approach 10\% precision will need to more carefully consider and estimate the
strength of the virtual $E1$ breakup.
In addition, a calculation was performed using a
coupled-channel code to estimate the effect of $E2$ excitations to the unbound
3$^+_1$ level, the level closest to the neutron threshold. It was found that
the cross section for this excitation was a factor of 1000 less likely than to
the first excited 1$^+$ level. It is therefore unlikely then that two-step processes to known levels would significantly contribute to our deduced $B(E2)$
value.

Last, a possible mechanism for the population of the first 1$^+$ state in
$^8$Li is the excitation from the non-resonant continuum in the Coulomb
excitation process. The possible magnitude of this contribution is currently
unknown. A future reaction theory calculation using, for example, the Extended
Continuum Discretized Coupled Channels method
\cite{Nunes1996,Summers2006,Summers2014}, would be able to estimate such a contribution.

\section{Discussion}
\label{sec:discussion}

\newcolumntype{a}{D{!}{}{-1}}

\begin{table*}[t]
  \caption{Experimental ground-state $E2$ moments~\cite{stone2016:e2-moments}, for $\isotope[8]{Li}$ and selected neighboring nuclides,
    and transition strengths to the first excited rotational band
    member~\cite{npa2002:005-007,npa2004:008-010,henderson2019:7be-coulex},
    where available.
    The rotational intrinsic quadrupole moment extracted from the experimental
    ground state quadrupole moment is given, along with the corresponding
    expected rotational $E2$ strength.}
  \label{tab:expt-rot}
  \newcommand{\missing}{\multicolumn{1}{c}{---}}
  \begin{center}
    \begin{ruledtabular}
      \begin{tabular}{llalardd}
        &\multicolumn{4}{c}{Experiment}
        &\multicolumn{3}{c}{Rotational}
        \\
        \cline{2-5}\cline{6-8}
        \multicolumn{1}{c}{Nuclide}
        & $J^P_{\text{gs}}$ & \multicolumn{1}{c}{$Q$ ($\fm^2$)} & $J^P_{\text{ex}}$ &\multicolumn{1}{c}{$B(E2\uparrow)$ ($e^2\fm^4$)}
        & $K$ & \multicolumn{1}{c}{$Q_0$ ($\fm^2$)}  &\multicolumn{1}{c}{$B(E2\uparrow)$ ($e^2\fm^4$)} \\
        \hline
        $\isotope[7]{Li}$ &
        $3/2^-$ & -4!.00(3) & $1/2^-$ & 8!.3(5) &
        $1/2$ & \approx +20 & \approx 8.0
        \\
        $\isotope[7]{Be}$ &
        $3/2^-$ & \approx-6!.8\footnotemark[1] & $1/2^-$ & 26!(6)(3)\footnotemark[2] &
        $1/2$ & \approx +34 & \approx 23
        \\
        $\isotope[8]{Li}$ &
        $2^+$ & +3!.14(2) & $3^+$ & \multicolumn{1}{c}{---} &
        $2$& \approx +11& \approx 6.0
        \\
        &
        & & $1+$ & 19!^{+7}_{-6}(2)\footnotemark[3]&
        $2\rightarrow1$& & \lesssim 2
        \\
        $\isotope[8]{B}$ &
        $2^+$ & +6!.34(14) & $3^+$ & \multicolumn{1}{c}{---} &
        $2$& \approx +22& \approx 25
        \\
        $\isotope[9]{Be}$ &
        $3/2^-$ & +5!.29(4) & $5/2^-$ & 42!(3)&
        $3/2$ & \approx +27 & \approx 36
        \\
      \end{tabular}
    \end{ruledtabular}
\end{center}
\scriptsize \raggedright
\footnotemark[1] Estimated from mirror nuclide quadrupole moment via \textit{ab initio} calculations~\cite{caprio2021:emratio}.
\\
\footnotemark[2] Indicated experimental uncertainties, from Ref.~\cite{henderson2019:7be-coulex}, are statistical and systematic, respectively.
\\
\footnotemark[3] Present work.
\end{table*}

The present measured $B(E2;2^+\rightarrow1^+)=19^{+7}_{-6}(2)\,e^2\fm^4$
 in $\isotope[8]{Li}$, while still enhanced relative to the
Weisskopf single-particle strength ($\approx0.95\,e^2\fm^4$ for $A=8$), is
reduced relative to the prior reported
$55(15)\,e^2\fm^4$~\cite{brown1991:8li-coulex}.  The present value is now well
within the range of $E2$ strengths observed in neighboring $A=7$ and $9$
nuclides, which reach $\approx 42 \,e^2\fm^4$ in $\isotope[9]{Be}$.

Nonetheless, the measured strength is still difficult to accommodate within
structural understanding of this nuclide.  The enhanced $E2$ strengths in
neighboring nuclides are readily understood in terms of collective rotational
enhancement, which cannot so simply explain an enhanced $2^+\rightarrow 1^+$
strength in $\isotope[8]{Li}$~\cite{caprio2022:8li-trans}.  Even assuming the $2^+$ ground state to be the
band head of a $K=2$ rotational band, there would be no $1^+$ band member.
\textit{Ab initio} theory, while successfully reproducing
enhanced rotational transitions in neighboring nuclides, does not predict
enhancement of the $2^+\rightarrow1^+$ strength in $\isotope[8]{Li}$~\cite{caprio2022:8li-trans}.

The deformation and the corresponding $E2$ enhancement in $\isotope[8]{Li}$ and its neighbors
is understood to arise from cluster
structure~\cite{vonoertzen1996:be-molecular,*vonoertzen1997:be-alpha-rotational,freer2007:cluster-structures,kanadaenyo2012:amd-cluster}.
In a cluster molecular picture for these nuclei, these nuclei are cluster
dimers, with possible additional nucleons occupying molecular orbitals around
this dimer core.  Thus, $\isotope[7]{Li}$ may be viewed as $\alpha+t$,
$\isotope[8]{Be}$ as $\alpha+\alpha$, $\isotope[9]{Be}$ as $\alpha+\alpha+n$,
\textit{etc.}  It is thus reasonable to expect axially-symmetric rotation.  In a
rotational strong coupling
picture~\cite{bohr1998:v1,rowe2010:collective-motion}, any additional nucleons
are taken to be coupled to the axially-symmetric rotational core.

Directly comparing the different $E2$ strengths in $\isotope[8]{Li}$ and its
neighbors is meaningless without taking into account the geometric factors
arising from rotational motion.  Recall that axially-symmetric rotation gives
rise to simple relations among all $E2$ matrix elements within a
band~\cite{alaga1955:branching,bohr1998:v1,rowe2010:collective-motion,maris2015:berotor2,*maris2019:berotor2-ERRATUM}.
In particular, all states within a rotational band share the same rotational
intrinsic state $\tket{\phi_K}$ in the body-fixed frame, and the spectroscopic
quadrupole moments $Q(J)$ and reduced transition probabilities
$B(E2;J_i\rightarrow J_f)$ are all obtained in terms of a single intrinsic $E2$
matrix element, or intrinsic quadrupole moment $Q_0$.\footnote{Namely,
  \begin{math}
    Q(J)=[3K^2-J(J+1)]/[(J+1)(2J+3)]Q_0,
  \end{math}
    and
    \begin{math}
      B(E2;J_i\rightarrow J_f)=
      (16\pi/5)^{-1}
      \tcg{J_i}{K}{2}{0}{J_f}{K}^2(eQ_0)^2,
    \end{math} where
    \begin{math}
      eQ_0\equiv(16\pi/5)^{1/2}\tme{\phi_K}{Q_{2,0}}{\phi_K}.
    \end{math}}  The rotational $E2$ strengths are thus tied to the ground state
quadrupole moments, which are precisely-measured for many of these
nuclei~\cite{stone2016:e2-moments}.

In $\isotope[7]{Li}$, the close-lying $3/2^-$ ground state and $1/2^-$ excited
state are interpreted as members of a $K=1/2$ rotational band, where the energy
order is inverted due to Coriolis staggering~\cite{rowe2010:collective-motion}
(see, \textit{e.g.}, Fig.~3 of Ref.~\cite{caprio2020:bebands}).  For
$\isotope[7]{Li}$ the measured
$Q(3/2^-)=-4.00(3)\,\fm^2$~\cite{stone2016:e2-moments} yields an intrinsic
quadrupole moment of $Q_0\approx+20\,\fm^2$, and thus a rotational prediction
$B(E2; 3/2^-\rightarrow 1/2^-)\approx8.0\,e^2\fm^4$, consistent with the
experimental $8.3(5)\,e^2\fm^4$~\cite{npa2002:005-007}.  (This and subsequent
rotational comparisons are summarized in Table~\ref{tab:expt-rot}.)

For $\isotope[7]{Be}$, the ground state quadrupole moment is not experimentally
known.  However, in a cluster picture, $\isotope[7]{Be}$ is obtained from
$\isotope[7]{Li}$ by replacing the triton ($\isotope[3]{H}$) cluster with a
$\isotope[3]{He}$ cluster.  In the limit of well-separated point clusters,
\textit{i.e.}, a ``ball-and-stick model'', this substitution gives
$Q(\isotope[7]{Be})/Q(\isotope[7]{Li})=50/34\approx1.5$~\cite{caprio2021:emratio},
yielding $Q_0\approx+29\,\fm^2$, while \textit{ab initio} predictions give a
somewhat larger ratio of
$\approx1.7$~\cite{pastore2013:qmc-em-alt9,caprio2021:emratio}, yielding
$Q_0\approx+34\,\fm^2$.  Thus, in a rotational picture, we expect $B(E2;
3/2^-\rightarrow 1/2^-)\approx23\,e^2\fm^4$ for $\isotope[7]{Be}$, again
consistent with the experimental value
$26(6)(3)\,e^2\fm^4$~\cite{henderson2019:7be-coulex}.

In the heavier neighbor $\isotope[9]{Be}$, the $3/2^-$ ground state is
understood to be the band head of a $K=3/2$ rotational band, with the $5/2^-$
excited state as a band member (see, \textit{e.g.}, Fig.~1 of
Ref.~\cite{caprio2020:bebands}).  The measured
$Q(3/2^-)=+5.29(4)\,\fm^2$~\cite{stone2016:e2-moments} yields an intrinsic
quadrupole moment of $Q_0\approx+27\,\fm^2$, and thus a rotational prediction
$B(E2; 3/2^-\rightarrow 5/2^-)\approx36\,e^2\fm^4$, on the same order as the
experimental $42(3)\,e^2\fm^4$~\cite{npa2004:008-010}, albeit not to within
experimental uncertainties.

The ground state of $\isotope[8]{B}$, mirror nuclide to $\isotope[8]{Li}$, has
been interpreted as consisting of a proton coupled to a deformed
$\isotope[7]{Be}$ core.  A spatially-extended molecular orbital for the proton
has been proposed (from fermionic molecular dynamics calculations) to lead to
proton halo structure~\cite{henninger2015:8b-fmd} and also suggested to
contribute to the enhanced ground-state quadrupole
moment~\cite{minamisono1992:8b-quadrupole-beta-nmr,kitagawa1993:shell-8b-17f-quadrupole-moment-halo}.
Assuming the ground state to be a $K=2$ band head, the measured
$Q(2^+)=+6.34(14)\,\fm^2$~\cite{stone2016:e2-moments} yields an intrinsic
quadrupole moment of $Q_0\approx+22\,\fm^2$ (Table~\ref{tab:expt-rot}).

However, returning to $\isotope[8]{Li}$, the measured
$Q(2^+)=+3.14(2)\,\fm^2$~\cite{stone2016:e2-moments}, at about half that of
$\isotope[8]{B}$, yields an intrinsic quadrupole moment of only
$Q_0\approx+11\,\fm^2$, the lowest among the nuclides discussed thus far by
nearly a factor of $2$ (Table~\ref{tab:expt-rot}), making any $E2$ enhancement
harder to explain.  Such a lower intrinsic quadrupole moment, relative to the
mirror nuclide $\isotope[8]{B}$, is reasonable in light of the lower quadrupole
moment of the $\isotope[7]{Li}$ core and the replacement of the charged halo
proton by an uncharged halo neutron.

For the $2^+\rightarrow1^+$ transition, the $1^+$ first excited state in
$\isotope[8]{Li}$ cannot be a member of the $K=2$ ground state band.  Even if
this $1^+$ state is rotational in nature, it must rather be a band head in its
own right~--- of either a $K=1$ band or, conceivably, a $K=0$ band with negative
signature (and thus
$J=1,3,\dots$~\cite{bohr1998:v1,rowe2010:collective-motion}).  The transition to
this state is then interband transition, requiring a change in the rotational
intrinsic wave function.

In the cluster molecular orbital description, we may expect the valence neutron
to be in a $\pi$ orbital with $K=3/2$, as in the isotone
$\isotope[9]{Be}$~\cite{dellarocca2018:cluster-shell-model-part1-9be-9b}.  The
$K$ quantum number adds algebraically, so the $K=2$ ground state band is
obtained from aligned coupling of this neutron with the $K=1/2$
$\isotope[7]{Li}$ core, while the $1^+$ excited state is ostensibly the band
head of a $K=1$ band arising from the antialigned coupling.

Even if, as a generous upper limit, we were to take the interband intrinsic
matrix element $\tme{\phi_{K=1}}{Q_{2,-1}}{\phi_{K=2}}$ to be identical in size
to the diagonal intrinsic matrix element $\tme{\phi_{K=2}}{Q_{2,0}}{\phi_{K=2}}$
determining the ground state band's intrinsic quadrupole moment, the rotational
picture would give $B(E2;2^+\rightarrow
1^+)/[eQ(2^+)]^2\lesssim0.24$,\footnote{As a point of comparison, in the mirror
  nuclide $\isotope[8]{B}$, fermionic molecular dynamics calculations give
  $Q(2^+)\approx4.9\,\fm^2$ and $B(E2;2^+\rightarrow 1^+)\approx3.3\,e^2\fm^4$,
  and thus, as a ratio, $B(E2;2^+\rightarrow
  1^+)/[eQ(2^+)]^2\approx0.14$~\cite{henninger2015:8b-fmd}.}  or
$B(E2;2^+\rightarrow 1^+) \lesssim 2\,e^2\fm^4$.  Only under the less-motivated
assumption that the $1^+$ excited state is the band head of a negative-signature
$K=0$ band does a transition on the scale of the present measurement become
plausible.  Taking the interband intrinsic matrix element, now
$\tme{\phi_{K=0}}{Q_{2,-2}}{\phi_{K=2}}$, to again be of the same size as the
diagonal intrinsic matrix element gives $B(E2;2^+\rightarrow
1^+)/[eQ(2^+)]^2\lesssim0.98$, or $B(E2;2^+\rightarrow 1^+) \lesssim
10\,e^2\fm^4$, but is the less favored scenario.


\section{Summary}

We have performed a radioactive-beam Coulomb excitation experiment to remeasure
the $2^+ \rightarrow 1^+$ $E2$ strength in $\isotope[8]{Li}$, now making use of
particle-$\gamma$ coincidences.  Compared to the previously-reported
$B(E2;2^+\rightarrow 1^+)=55(14)\,e^2\fm^4$~\cite{brown1991:8li-coulex}, our
measured value of $19^{+7}_{-6}(2)\,e^2\fm^4$ is smaller by approximately a factor of
three, but remains difficult to accommodate within a theoretical
understanding of the nuclear structure.

The enhanced $E2$ strengths in the neighboring $A=7$ and $9$ nuclei are
naturally understood in terms of in-band rotational transitions, and we find
that they are well-described by \textit{ab initio} predictions.  However, there is no
simple way to explain an enhanced $2^+ \rightarrow 1^+$ transition in
$\isotope[8]{Li}$.  Such a transition is not naturally viewed as a rotational
in-band transition, and it is unnaturally large for an interband transition.

The magnitude of
possible contributions from the non-resonant continuum to the cross section, and
thus to the $E2$ strength extracted from experiment, is unknown.  Calculations
from reaction theory to estimate such contributions from the continuum could
clarify if they might explain the discrepancy between experiment and theory.
Other possible contributions to the cross section, from two-step Coulomb
excitation (or dipole polarization), involving virtual excitation of
higher-lying states above the neutron separation threshold, as well as from
indirect feeding involving direct excitation of these states, are estimated to
be insufficient to explain the discrepancy.  The use of particle-$\gamma$
coincidences in the present experiment eliminates any significant contribution
from $\isotope[8]{Li}$ $1^+ \rightarrow 2^+$ $\gamma$ rays that are produced in
the primary reaction, removing a possible experimental inflation of the $B(E2)$
value.  In conclusion, our studies confirm that there is no easy interpretation
for a large $B(E2)$ in $\isotope[8]{Li}$, and our experimental result reinforces
the tension between experiment and theory suggested by the previous work.

\section*{Acknowledgments}
We thank Colin V.~Coane, Jakub Herko, and Zhou Zhou for comments on the
manuscript and Filomena Nu\~nes for discussions on contributions from the non-resonant continuum.  This work was supported by the U.S.~National Science Foundation
under Grant Nos.~PHY 20-11890, PHY 17-13857, PHY 14-01343, and PHY 14-30152, and by the
U.S.~Department of Energy, Office of Science, under Award
Nos.~DE-FG02-95ER40934, DE-SC021027, and FRIB Theory Alliance award DE-SC0013617 and  sponsored in part by the National Nuclear Security Administration under the Stewardship Science Academic Alliance program through DOE Cooperative Agreement No. DE-NA000213.   This
research used computational resources of the University of Notre Dame Center for
Research Computing and of the National Energy Research Scientific Computing
Center (NERSC), a U.S.~Department of Energy, Office of Science, user facility
supported under Contract DE-AC02-05CH11231.
Figures in this work were produced using ROOT \cite{Brun1997}, \texttt{matplotlib} \cite{hunter2007:matplotlib}, and \texttt{seaborn} \cite{waskom2021:seaborn}.

\bibliographystyle{apsrev4-2}
%

\end{document}
%